\documentclass[aps, prl, twocolumn, superscriptaddress, floatfix]{revtex4}
\usepackage{natbib}
\usepackage{amsmath} % AMS Math Package
\usepackage{amssymb}	% Math symbols such as \mathbb
\usepackage{graphicx} % Allows for eps images
\usepackage{times}
\usepackage{color}

\begin{document}

%\begin{titlepage}
%This manuscript has been authored by UT-Battelle, LLC under Contract No. DE-AC05-00OR22725 with the U.S. Department of Energy. The United States Government retains and the publisher, by accepting the article for publication, acknowledges that the United States Government retains a non-exclusive, paid-up, irrevocable, world-wide license to publish or reproduce the published form of this manuscript, or allow others to do so, for United States Government purposes. The Department of Energy will provide public access to these results of federally sponsored research in accordance with the DOE Public Access Plan (http://energy.gov/downloads/doe-public-access-plan).
%\pagebreak[4]
%\end{titlepage}

\title{Extended exchange interactions stabilize long-period magnetic structures in Cr$_{1/3}$NbS$_2$}

\author{A. A. Aczel}
\altaffiliation{author to whom correspondences should be addressed: E-mail:[aczelaa@ornl.gov]}
\affiliation{Neutron Scattering Division, Oak Ridge National Laboratory, Oak Ridge, TN 37831, USA}
\author{L.M DeBeer-Schmitt}
\affiliation{Neutron Scattering Division, Oak Ridge National Laboratory, Oak Ridge, TN 37831, USA}
\author{T.J. Williams}
\affiliation{Neutron Scattering Division, Oak Ridge National Laboratory, Oak Ridge, TN 37831, USA}
\author{M.A. McGuire}
\affiliation{Materials Science and Technology Division, Oak Ridge National Laboratory, Oak Ridge, TN 37831, USA}
\author{N.J. Ghimire}
\altaffiliation{Materials Science Division, Argonne National Laboratory, 9700 South Cass Avenue, Lemont , IL 60439 USA}
\affiliation{Materials Science and Technology Division, Oak Ridge National Laboratory, Oak Ridge, TN 37831, USA}
\affiliation{Department of Materials Science and Engineering, University of Tennessee, Knoxville, TN 37996, USA}
\author{L. Li}
\affiliation{Department of Materials Science and Engineering, University of Tennessee, Knoxville, TN 37996, USA}
\author{D. Mandrus}
\affiliation{Materials Science and Technology Division, Oak Ridge National Laboratory, Oak Ridge, TN 37831, USA}
\affiliation{Department of Materials Science and Engineering, University of Tennessee, Knoxville, TN 37996, USA}

\date{\today}% It is always \today, today, but any date may be explicitly specified

\begin{abstract}
The topologically-protected, chiral soliton lattice is a unique state of matter offering intriguing functionality and it may serve as a robust platform for storing and transporting information in future spintronics devices. While the monoaxial chiral magnet Cr$_{1/3}$NbS$_2$ is known to host this exotic state in an applied magnetic field, its detailed microscopic origin has remained a matter of debate. Here we work towards addressing this open question by measuring the spin wave spectrum of Cr$_{1/3}$NbS$_2$ over the entire Brillouin zone with inelastic neutron scattering. The well-defined spin wave modes allow us to determine the values of several microscopic interactions for this system. The experimental data is well-explained by a Heisenberg Hamiltonian with exchange constants up to third nearest neighbor and an easy plane magnetocrystalline anisotropy term. Our work shows that both the second and third nearest neighbor exchange interactions contribute to the formation of the helimagnetic and chiral soliton lattice states in this robust three-dimensional magnet. 
\end{abstract}

\maketitle

Chiral objects, which are non-superimposable on their mirror images, are ubiquitous in nature. These objects appear in a wide variety of forms, including molecules, biological systems, and magnetic materials. They provide a great deal of functionality \cite{07_wagniere} and therefore are poised to be the catalyst to innovative technologies. 

The crystal structures of chiral magnets lack both inversion centers and mirror planes and therefore the antisymmetric Dzyaloshinskii-Moriya interaction (DMI) is symmetry-allowed. Competition between Heisenberg exchange, the DMI, magnetocrystalline anisotropy and an applied magnetic field can lead to interesting, non-trivial spin textures such as the skrymion lattice (SkX) originally proposed by Bogdanov and Hubert \cite{94_bogdanov}. This topologically-protected phase consists of an array of magnetic vortices with particular spin windings and it has now been established as an accessible phase of matter in the temperature-magnetic field phase diagram of a wide variety of cubic, chiral helimagnets \cite{09_muhlbauer, 10_munzer, 10_yu, 10_yu_2}. The SkX state is particularly attractive for many potential spintronics applications, including skyrmion-based race track memory, logic gates, and transistors \cite{16_kang}, due to the mesoscale length scale (5-100~nm) of an individual skyrmion \cite{13_nagaosa} and the small current densities (~10$^{6}$~A/m$^2$ for MnSi \cite{10_jonietz}) required to manipulate the lattice through the spin-torque effect.  

Magnetocrystalline anisotropy plays a minor role in the stability of the skyrmion lattice state in the phase diagrams of MnSi and isostructural helimagnets \cite{12_bauer} due to its weak contribution to the Hamiltonian of cubic systems. This feature is advantageous when looking for skyrmion hosts because the applied field direction is not critical for the detection of the skyrmion state. On the other hand, the stronger magnetocrystalline anisotropy inherent to lower-symmetry crystal systems can be beneficial for enhancing stability of the SkX state in certain cases \cite{10_butenko, 12_karhu, 16_vousden} and it can also lead to entirely unexpected magnetic phases. For example, the chiral soliton lattice (CSL) was observed recently in the monoaxial chiral helimagnet Cr$_{1/3}$NbS$_2$, which is a topologically-protected phase consisting of ferromagnetic regions partitioned by helical domain walls \cite{12_togawa}. This state arises when a static magnetic field is applied perpendicular to the helical axis and its size and shape can be controlled by varying the field. The high tunability, robust nature, and other unique properties of the CSL phase are intriguing in the context of spintronics devices with information storage or transportation capabilities \cite{09_borisov, 10_kishine, 17_kim}.   

The detailed microscopic origin of the CSL phase remains unknown due to the absence of a robust Hamiltonian for Cr$_{1/3}$NbS$_2$. For this reason, the magnetic properties have been most commonly studied by the chiral sine-Gordon model \cite{07_kishine, 16_shinoazaki}. In its simplest form, this model consists of Cr spins on a 1D lattice that are only coupled by nearest neighbor FM exchange $J_\parallel$ and a DMI term with $\vec{D}$ pointing along the chain direction. A helical magnetic ground state arises in zero field due to the competition between these two terms, with the turn angle $\alpha$ between spins at adjacent sites given by $\alpha$~$=$~$\arctan(D/J_\parallel)$. This simple model can be used to calculate $M$ vs $T$ \cite{07_kishine}, $M$ vs $H$ \cite{09_kousaka}, and the field-dependence of the electron spin resonance spectra \cite{09_kishine} that are characteristic of the CSL phase and reproduces the corresponding experimental data well for appropriate choices of $D$ and $J_\parallel$ \cite{14_chapman, 15_yoshizawa}. Numerical simulations with the chiral sine-Gordon model and an extended version including nearest neighbor intraplane coupling $J_\perp$ have also been performed \cite{16_shinoazaki}. These calculations enabled the authors to estimate $J_\parallel$, $J_\perp$, and $D$ and they found that $J_\perp$~$>>$~$J_\parallel$, which suggests that Cr$_{1/3}$NbS$_2$ is a quasi-two-dimensional (2D) magnet. Notably, there is significant variance in the values of $J_\parallel$ and $D$ reported in these three works. This observation calls into question the simple assumption of the chiral sine-Gordon model for Cr$_{1/3}$NbS$_2$, namely $\alpha$~$=$~$\arctan(D/J_\parallel)$. Furthermore, recent first principles work \cite{16_mankovsky} calculated the exchange constants for Cr$_{1/3}$NbS$_2$ and found that $J_\parallel$~$>$~$J_\perp$, a scenario more consistent with a robust three-dimensional (3D) magnet. These two discrepancies have remained unresolved in the literature. 

Cr$_{1/3}$NbS$_2$ is one member of a large family of metallic, monoaxial chiral magnets that have received little attention overall \cite{80_parkin}. Assuming that a local moment model is applicable for this family of materials, careful parameterization of the Hamiltonian can be achieved by measuring spin wave dispersions along different crystallographic directions over the entire Brillouin zone with inelastic neutron scattering (INS). In this letter, we use INS to measure the magnetic excitation spectrum of Cr$_{1/3}$NbS$_2$ in the helimagnetic state. We find that the spin waves are well-defined throughout the majority of the Brillouin zone, and therefore we are able to determine the Heisenberg exchange constants and magnetocrystalline anisotropy using linear spin wave theory. Our work presents the first step towards developing a complete local moment model that can explain the microscopic origin of the helimagnetic and CSL states observed in this material. This model may provide insight on how this material can be modified to optimize the CSL state for future applications or to realize other interesting spin textures such as the SkX state. We note that there is a great opportunity to tune Cr$_{1/3}$NbS$_2$ through uniaxial strain \cite{17_fobes} or chemical substitution \cite{80_parkin}.

Cr$_{1/3}$NbS$_2$ single crystals were grown at the University of Tennessee by chemical vapor transport with iodine serving as the transport agent \cite{13_ghimire}; additional growth details are provided in Ref.~\cite{83_miyadai}. Magnetization data was collected with a Magnetic Property Measurement System at Oak Ridge National Laboratory (ORNL) to verify that the crystals showed evidence for the CSL state in modest magnetic fields. 

INS experiments were performed on the direct-geometry time-of-flight chopper spectrometer SEQUOIA at the Spallation Neutron Source at ORNL. An array of 20 single crystals with a total mass of $\sim$1~g was co-aligned with a 5$^\circ$ mosaic spread and loaded with the (H0L) plane horizontal in a 8 K base temperature closed cycle refrigerator for this experiment. Spectra were constructed by performing 18~minute runs at several different sample orientations rotating about the [-1 2 0] direction and then combining these runs into a single data file for a given set of instrument conditions. We used an incident energy $E_i$~$=$~50 meV, a Fermi chopper frequency of 180~Hz, and a $T_{0}$ chopper frequency of 90~Hz, providing an energy resolution at the elastic line of approximately 2.7~meV. Additional INS measurements with much finer energy resolution were performed on the cold triple axis spectrometer CTAX at the High Flux Isotope Reactor at ORNL. The same co-aligned array of crystals was loaded into a cryostat with a base temperature of 1.5~K for this experiment. The measurements were performed using a fixed final energy $E_f$~$=$~5 meV and collimation settings of guide-open-80'-open, which generated an energy resolution at the elastic line of 0.3~meV. This fine energy resolution enabled us to put stringent constraints on an energy gap associated with the spin wave spectrum. No magnetic field was applied during the INS experiments. 

\begin{figure}
\centering
\scalebox{0.24}{\includegraphics{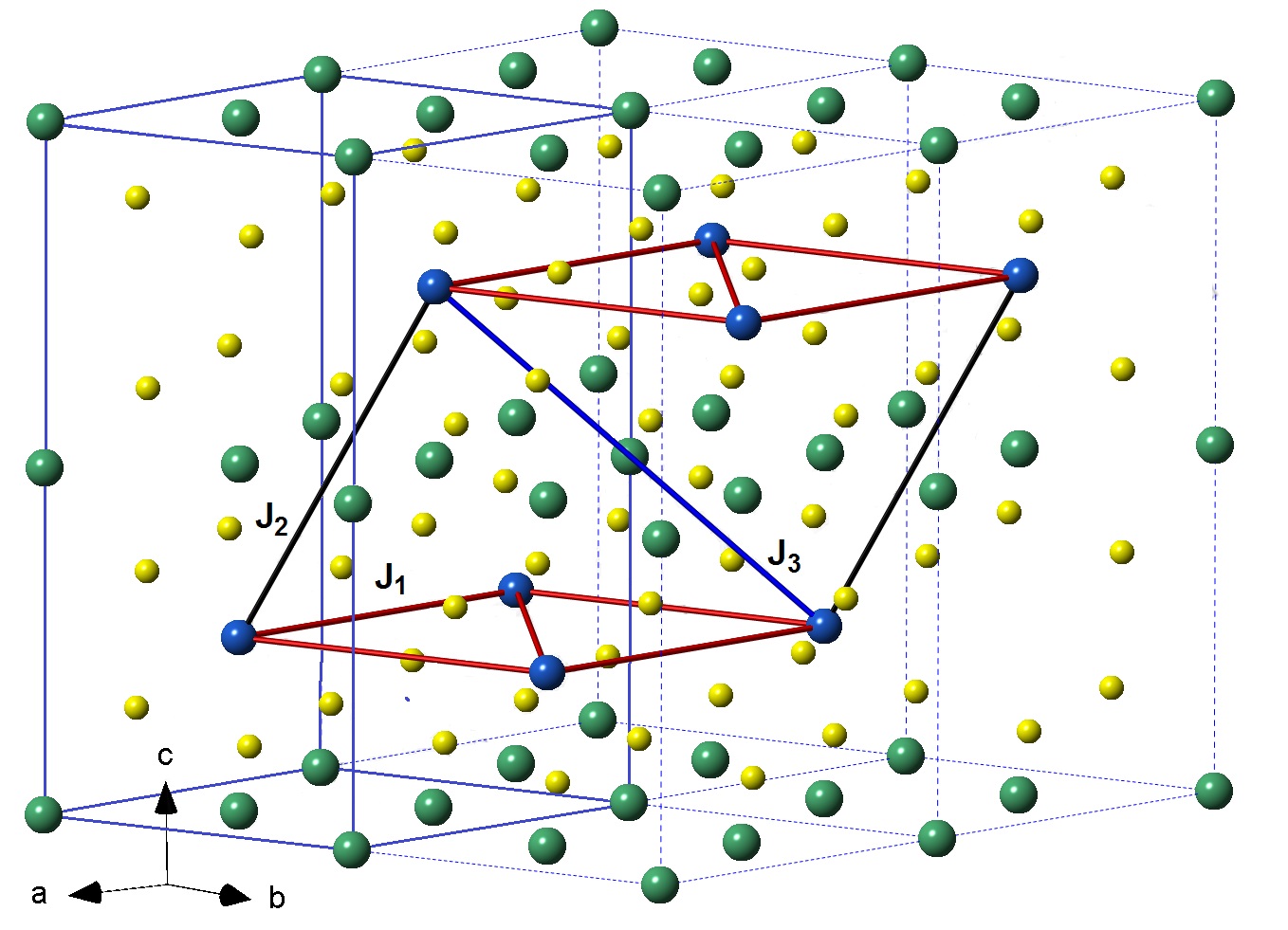}}
\caption{\label{Fig1} (color online) A schematic of the Cr$_{1/3}$NbS$_2$ crystal structure, depicting the Cr atoms with blue spheres. The three Heisenberg exchange pathways considered in this work are $J_1$, $J_2$, and $J_3$. }
\end{figure}

\begin{figure}
\centering
\scalebox{0.45}{\includegraphics{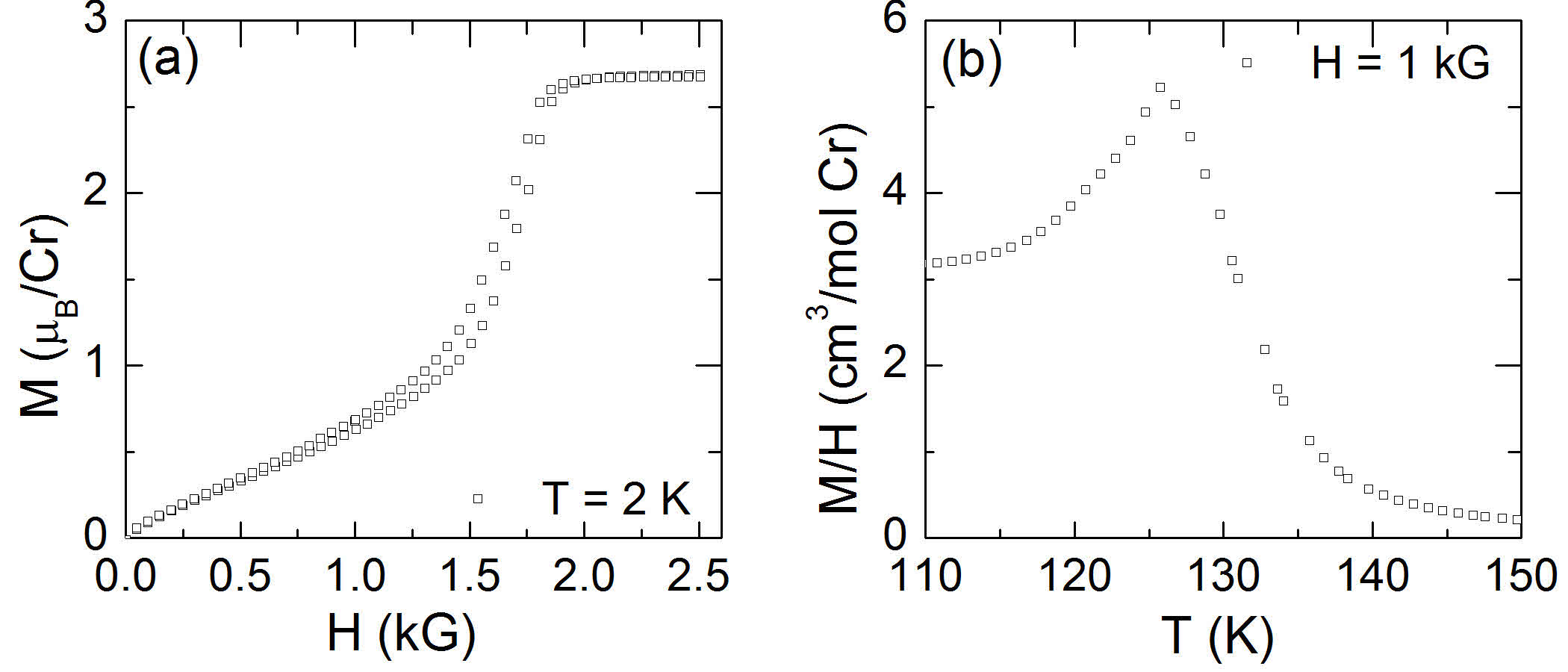}}
\caption{\label{Fig2} (color online) (a) $M$ vs $H$ for a representative Cr$_{1/3}$NbS$_2$ single crystal investigated by INS in this work. (b) $M/H$ vs $T$ for the same Cr$_{1/3}$NbS$_2$ single crystal.}
\end{figure}

\begin{figure*}
\centering
\scalebox{0.55}{\includegraphics{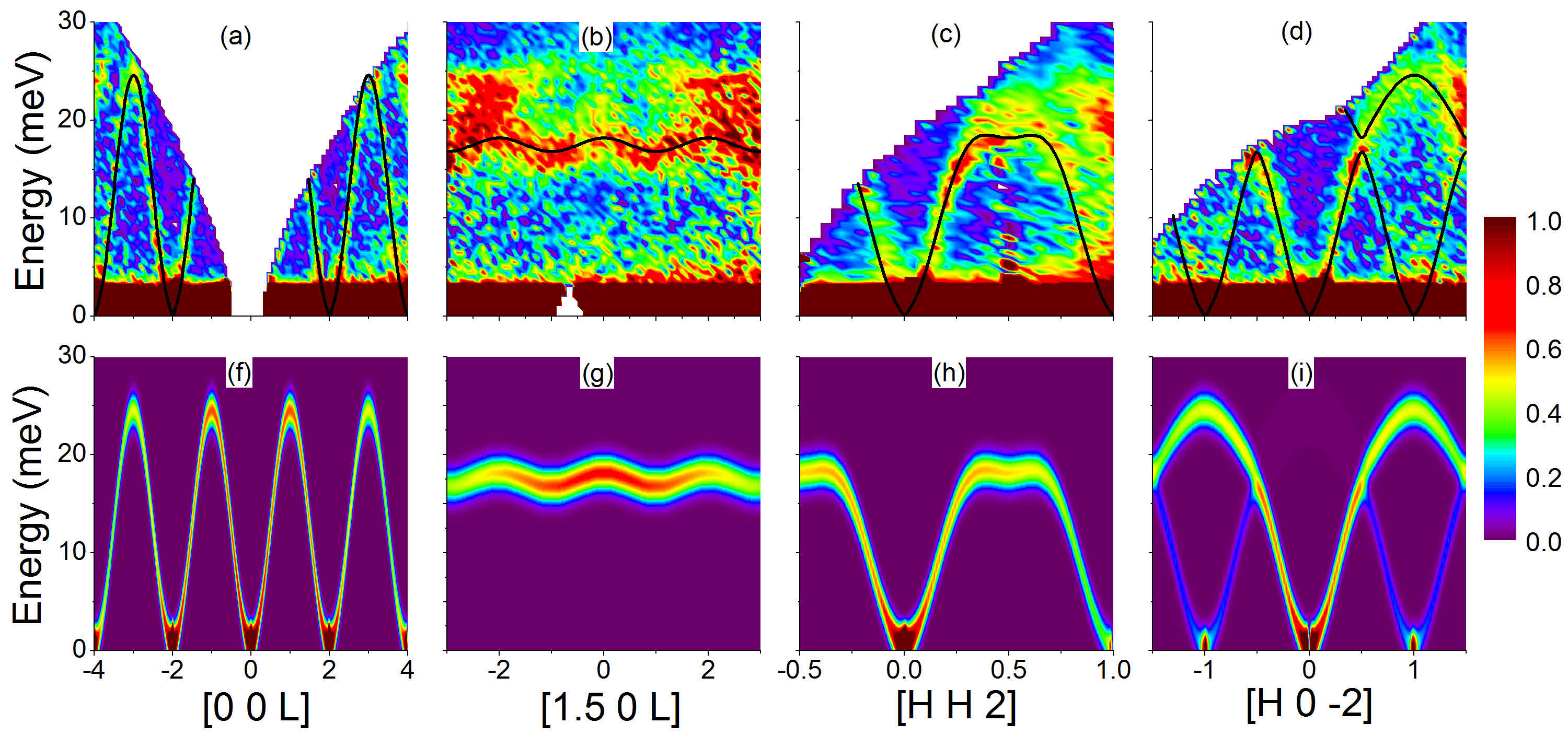}}
\caption{\label{Fig3} (color online) Energy-momentum color contour plots from SEQUOIA at $T$~$=$~8~K with $E_i$~$=$~50~meV: (a) energy transfer $E$ vs [0 0 L], (b) $E$ vs [1.5 0 L], (c) $E$ vs [H H 2], and (d) $E$ vs [H 0 -2]. Integration ranges along the two perpendicular momentum axes were 0.15 reciprocal lattice units (rlu) for directions in the (HK0) plane and 0.3 rlu along [0 0 L]. The black curves superimposed on the data represent the dispersion relations calculated using the best fit parameters for the $J_1$-$J_2$-$J_3$-$A$ Heisenberg model. (e)-(h) Energy-momentum color contour simulations calculated with linear spin wave theory using the best fit parameters for the $J_1$-$J_2$-$J_3$-$A$ Heisenberg model .}
\end{figure*}

Cr$_{1/3}$NbS$_2$ crystallizes in the chiral space group $P 6_3 2 2$; the crystal structure is illustrated in Fig.~\ref{Fig1}. The Cr atoms occupy the 2c Wyckoff site exclusively in the ideal structure. This arrangement results in AB stacking along the c-axis and a $\sqrt{3} \times \sqrt{3}$ superstructure within the ab-plane with respect to the unit cell of NbS$_2$. Cr$_{1/3}$NbS$_2$ has a helimagnetic ground state with a pitch of 47(1)~nm that onsets below $T_c$~$=$~127~K \cite{83_miyadai, 12_togawa}. The Cr moments are ferromagnetically-coupled in the ab-plane and the magnetic helix propagates along the c-axis. This system is very susceptible to Cr disorder since a fraction of the 2b and 2d Wyckoff sites may also be occupied instead of their 2c Wyckoff site counterparts. A significant amount of Cr disorder can have profound consequences on the material properties by modifying both the crystal and magnetic structures, ultimately leading to a centrosymmetric crystal structure hosting a ferromagnetic ground state with a $T_c$~$=$~88~K \cite{15_dyadkin}. 

Cr$_{1/3}$NbS$_2$ crystals with the chiral structure can be identified by performing magnetization measurements. Figure~\ref{Fig2}(a) depicts $M$ vs $H$ data with $\vec{H}$~$\perp$~$\hat{c}$ at 2 K for one of the crystals in our INS mosaic. The magnetization reaches a saturation value of 2.7~$\mu_B$ for $H$~$=$~2~kG, which is just under the expected value of 3~$\mu_B$ for localized $S$~$=$~3/2 magnetic moments. The general shape and hysteresis of the $M$ vs $H$ curve are consistent with the development of a chiral soliton lattice in finite magnetic fields \cite{07_kishine, 09_kousaka}. Figure~\ref{Fig2}(b) presents $M/H$ vs $T$ data at 1 kG with $\vec{H}$~$\perp$~$\hat{c}$. The characteristic cusp expected at $T_c$ when cooling into the CSL state is clearly visible \cite{07_kishine, 09_kousaka} and the value of $T_c$ agrees well with expectations for the chiral crystal structure. These combined magnetization measurements therefore illustrate that our representative Cr$_{1/3}$NbS$_2$ single crystal has the desired $P 6_3 2 2$ structure. We repeated these measurements on other crystals from our mosaic and obtained identical results. 

\begin{figure}
\centering
\scalebox{0.34}{\includegraphics{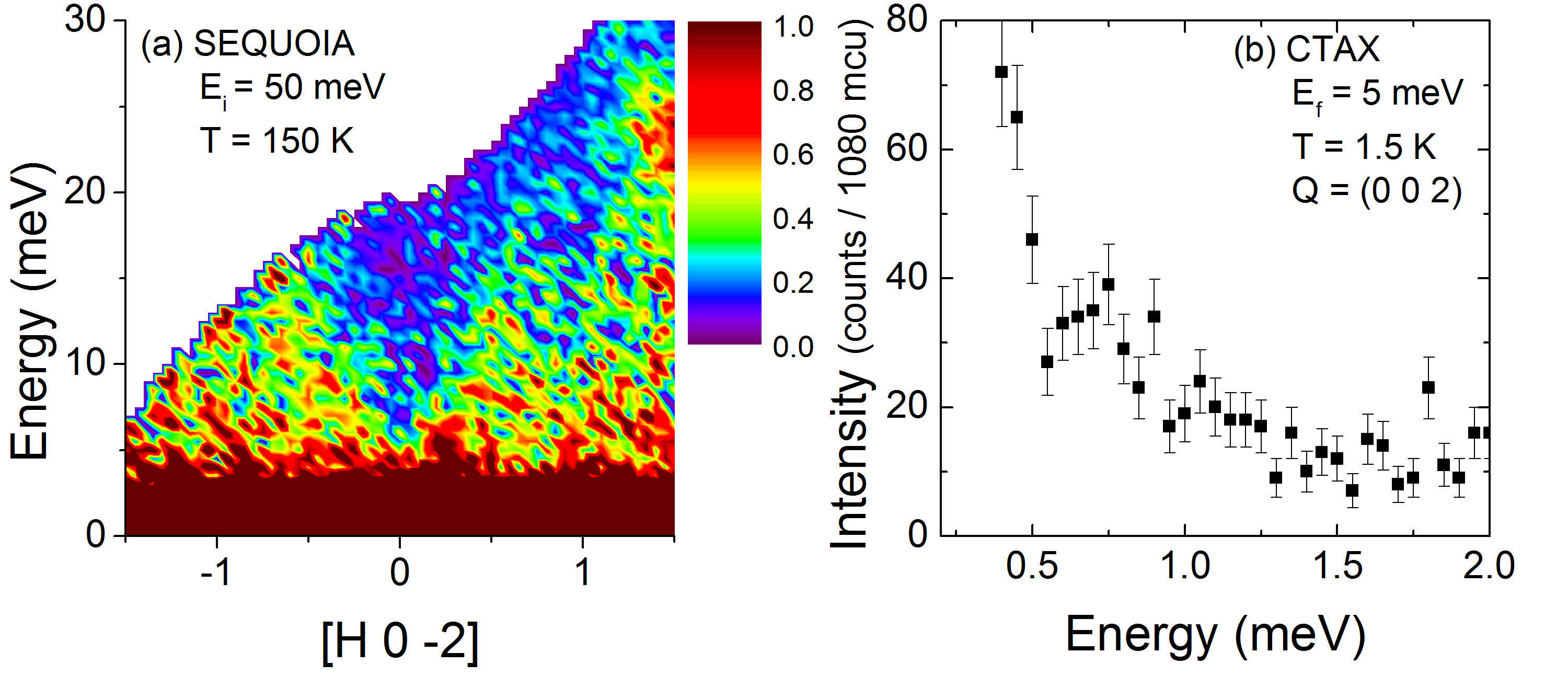}}
\caption{\label{Fig4} (color online) (a) Energy-momentum color contour plot from SEQUOIA for $E$ vs [H 0 -2] with $T$~$=$~150~K, $E_i$~$=$~50~meV, and integration ranges [0 0 L]~$=$~[-2.15, -1.85] rlu and [H -2H 0]~$=$~[-0.075, 0.075]~rlu. (b) A constant-$Q$ scan from CTAX at the Brillouin zone center (0 0 2) with $T$~$=$~1.5~K and $E_f$~$=$~5~meV. Note that 1 monitor count unit (mcu) equals 1 s when $E_i$~$=$~5~meV.   }
\end{figure}

Our time-of-flight neutron spectroscopy measurements were particularly useful for mapping out the magnetic dispersions and dynamical structure factors over the entire Brillouin zone. Representative energy-momentum color contour plots of the INS spectra at $T$~$=$~8~K along specific high-symmetry crystallographic directions are presented in Fig.~\ref{Fig3}(a)-(d). Well-defined modes appear to originate from the Brillouin zone centers (i.e. $\Gamma$ points) expected for ferromagnetic spin wave theory and they disappear above $T_c$~$=$~127~K as shown in Fig.~\ref{Fig4}(a) and the Supplementary Material (SM), which is indicative of a spin-wave origin. The sharp excitations observed throughout the majority of the Brillouin zone, combined with a saturation moment reduced only slightly from expectations for localized magnetism, are consistent with a metallic ferromagnet in the strong moment limit and therefore a Heisenberg model should describe the spin dynamics well. Complementary cold triple axis data, in the form of a constant $Q$-scan at the (002) zone center, is shown in Fig.~\ref{Fig4}(b). There is no indication of a magnon mode in this data, and therefore the spin wave spectrum is gapless within the 0.3~meV energy resolution of CTAX. This result is consistent with a magnetically-ordered system characterized by easy plane anisotropy, which was initially revealed by magnetization measurements \cite{83_miyadai}. 

A minimal Hamiltonian that can adequately explain the long-range helimagnetic ordering of Cr$_{1/3}$NbS$_2$ includes two Heisenberg exchange parameters, $J_1$ and $J_2$, an easy plane anisotropy term $A$, and a DMI contribution $D$ \cite{83_miyadai}. $J_1$ couples nearest neighbor (NN) Cr moments in the ab-plane, while $J_2$ and $D$ represent the NN interplane interactions. We neglect $D$ here because our INS experiments cannot resolve the momentum shift away from commensurate reciprocal space points of the magnetic Bragg peaks or of the spin wave dispersion minima that a non-zero $D$ should generate \cite{16_sato}. Also, we observe no evidence for the helimagnon bands expected to arise for non-zero $D$ \cite{10_janoschek}. We used the {\it SpinW} software \cite{15_toth} to compute the spin wave dispersions and dynamical structure factors with linear spin wave theory for this simple $J_1$-$J_2$-$A$ Heisenberg model. Our magnetic ground state consists of $S$~$=$~3/2 Cr moments ferromagnetically-aligned along the crystallographic a-axis. The Hamiltonian parameters were determined by using {\it SpinW}'s swarm particle optimization routine \cite{95_kennedy} to minimize the sum of the squared residuals obtained from comparing the experimental and model-calculated dispersion relations. The experimental dispersion relations were quantified by fitting a series of SEQUOIA constant-$Q$ and constant-$E$ cuts to Gaussian functions (plus a sloping background) along the four high-symmetry directions shown in Fig.~\ref{Fig3}. Representative constant-$Q$ cuts with the Gaussian fits superimposed are shown in the SM. This fitting routine was run to convergence 50 times and the interaction parameters reported in this work correspond to the average values, while the errors correspond to the standard deviations of the fit parameters. 

The best fit parameters for this model are $J_1$~$=$~-0.51(2)~meV, $J_2$~$=$~-1.22(1)~meV, and $A$~$=$~0.95(9)~meV. The calculated scattering intensity using these parameters is presented along several high-symmetry directions in the SM. Note that negative values for the exchange parameters correspond to ferromagnetic interactions, while the positive $A$ value represents an easy plane magnetocrystalline anisotropy. Both the Cr$^{3+}$ magnetic form factor and energy broadening were included in these calculations; the latter was accomplished by convolving the dynamical structure factor with a Gaussian function. The full-width-half-maximum of this function was fixed to 2.7~meV, which corresponds to the instrumental energy resolution at the elastic line for the $E_i$~$=$~50~meV data set. There are clear deviations between the calculation and the INS data for this simple model. More specifically, a large gap appears between the low and high energy modes along the [H 0 -2] direction that is not present in the experimental data and the dispersion along the [1.5 0 L] direction is significantly overestimated by the model. 

In an effort to improve the overall agreement between theory and experiment, we added a second interplane Heisenberg term to our model, $J_3$. The $J_1$-$J_3$ exchange paths are shown in Fig.~\ref{Fig1}. We used the same fitting procedure described above to obtain the Hamiltonian parameters in this case, and the calculated scattering intensity using the best fit parameters is presented along several high-symmetry directions in Fig.~\ref{Fig3}(e)-(h). We obtain excellent agreement between the INS data and the calculation using this model for the parameter set presented in Table~I. While there is extra scattering in the data along some of the high-symmetry directions between energy transfers of 20-25~meV that is not accounted for by the calculation, it likely arises from phonon scattering associated with the Al sample can since the spectral weight appears to increase with $Q$ and persists above $T_c$.  

Our final parameterization of the Hamiltonian for Cr$_{1/3}$NbS$_2$ implies that the RKKY mechanism is mediating the exchange interactions \cite{70_koehler, 16_williams}, as the long-range couplings that we find are too strong to have a superexchange origin. Notably, the NN and 2NN superexchange pathways consist of two S atoms while the 3NN superexchange pathway is even longer and mediated by three S atoms. Our best fit Hamiltonian parameters are also inconsistent with some of the assumptions and conclusions made in earlier work on Cr$_{1/3}$NbS$_2$. We find that the expression for the helical turn angle $\alpha$~$=$~$\arctan(D/J_\parallel)$ obtained from the chiral sine-Gordon model is not appropriate here since the Heisenberg exchange interactions between magnetic ions in adjacent ab-planes extend beyond NN, which contradicts the simple assumption made in this model. We also find that Cr$_{1/3}$NbS$_2$ does not consist of weakly-coupled Cr planes, as this would have led to a very weak dispersion along the [00L] direction \cite{15_williams} that is inconsistent with our INS data. Instead, Cr$_{1/3}$NbS$_2$ is a robust 3D magnet due to the similar values of the intraplane exchange $J_1$ and the interplane exchanges $J_2$ and $J_3$. These strong 3D interactions likely work together in a co-operative fashion to produce the high ordering temperature of $T_c$~$=$~127~K. Our findings are in excellent agreement with recent work reporting 3D Heisenberg critical exponents for Cr$_{1/3}$NbS$_2$ \cite{17_clements}.  

\begin{center}
\begin{table}[htb]
\caption{Microscopic parameters determined by fitting the inelastic neutron scattering data to the extended Heisenberg model described in the main text. Bonds are listed in the form ($x$, $y$, $z$), where $x$, $y$, and $z$ are fractions of the direct lattice vectors $\vec{a}$, $\vec{b}$, and $\vec{c}$.} 

\begin{tabular}{| c | c | c | c | c |}
\hline 
Exchange & No. neighbors & Bond & Distance (\AA) & Value (meV) \\  \hline
$J_1$ & 6 & (1, 0, 0) & 5.74 & -0.37(1) \\
$J_2$ & 6 & (1/3, -1/3, 1/2) & 6.90 & -1.02(2) \\ 
$J_3$ & 6 & (2/3, -2/3, 1/2) & 8.97 & -0.28(2) \\ 
$A$ & - & - & - & 0.8(1) \\ \hline
\end{tabular}

\end{table}
\end{center}

It is interesting to compare our INS results to recent theoretical work that determined the Hamiltonian parameters for Cr$_{1/3}$NbS$_2$ using first principles calculations \cite{16_mankovsky}. Their Heisenberg exchange constants, reported as $J_{ij} S^2$, reveal that the first three NN terms in Cr$_{1/3}$NbS$_2$ are dominant and the 2NN term provides the largest contribution to the Hamiltonian. These qualitative findings and the magnitudes reported for these parameters are in excellent agreement with our results presented here. The first principles work also finds an easy plane magnetocrystalline anisotropy on the order of 0.1~meV, which agrees well with an earlier experimental estimate of 0.4 meV using magnetization data \cite{83_miyadai} and our value obtained with INS. Interestingly, the first principles work determines values for interplane DMIs that we were unable to resolve in our work. They find that the DMIs corresponding to several different Cr-Cr interplane distances are comparable. This result implies that the microscopic details leading to the establishment of the long-period helimagnetic and soliton lattice states in Cr$_{1/3}$NbS$_2$ may be rather complicated, but this prediction has yet to be verified by experiment. 

In conclusion, we have presented the full spin wave spectrum of Cr$_{1/3}$NbS$_2$, as revealed by INS measurements. The spin wave dispersions and dynamical structure factors are explained well by a Hamiltonian with Heisenberg exchange constants up to 3rd nearest neighbor and an easy plane magnetocrystalline anisotropy. Our successful parameterization of the Cr$_{1/3}$NbS$_2$ INS data using a local moment model is important because it implies that this technique can be used for determining the exchange constants and magnetocrystalline anisotropy of other isostructural materials. Elucidating the microscopic origins of the long-period magnetic structures with effective Hamiltonians should be far more straightforward than establishing a direct link between this interesting magnetism and the band structure, which is a necessity for the weak itinerant magnet MnSi \cite{77_ishikawa, 11_boni}. Intercalated transition metal dichalcogenides are therefore excellent systems to pursue optimization of the CSL state for future applications or to search for CSL/SkX hosts via a systematic approach.  

\section{Supplementary Material}
See supplementary material for additional color contour plots of the $T$~$=$~150~K SEQUOIA data, representative constant-$Q$ cuts from the $T$~$=$~8~K SEQUOIA data, and spin wave simulations for the $J_1$-$J_2$-$A$ Heisenberg Hamiltonian described in the main text. 

\section{Acknowledgments}
We thank M.B. Stone for technical support. MAM acknowledges support from the US Department of Energy, Office of Science, Basic Energy Sciences, Materials Sciences and Engineering Division. A portion of this research used resources at the Spallation Neutron Source and the High Flux Isotope Reactor, which are DOE Office of Science User Facilities operated by Oak Ridge National Laboratory.

\end{document}